# Attention module improves both performance and interpretability of 4D fMRI decoding neural network

Zhoufan Jiang[1], Yanming Wang[1], ChenWei Shi[1], Yueyang Wu[1], Rongjie Hu[1], Shishuo Chen[1], Sheng Hu[1], Xiaoxiao Wang[1,*], Bensheng Qiu[1,*]

1. Hefei National Lab for Physical Sciences at the Microscale and the Centers for Biomedical Engineering, University of Science and Technology of China, Hefei, China.

**\*Correspondence:**
Bensheng Qiu, Email: bqiu@ustc.edu.cn
Xiaoxiao Wang, Email: wang506@ustc.edu.cn

## Abstract

Decoding brain cognitive states from neuroimaging signals is an important topic in neuroscience. In recent years, deep neural networks (DNNs) have been recruited for multiple brain state decoding and achieved good performance. However, the open question of how to interpret the DNN black box remains unanswered. Capitalizing on advances in machine learning, we integrated attention modules into brain decoders to facilitate an in-depth interpretation of DNN channels. A 4D convolution operation was also included to extract temporo-spatial interaction within the fMRI signal. The experiments showed that the proposed model obtains a very high accuracy (97.4%) and outperforms previous researches on the 7 different task benchmarks from the Human Connectome Project (HCP) dataset. The visualization analysis further illustrated the hierarchical emergence of task-specific masks with depth. Finally, the model was retrained to regress individual traits within the HCP and to classify viewing images from the BOLD5000 dataset, respectively. Transfer learning also achieves good performance. A further visualization analysis shows that, after transfer learning, low-level attention masks remained similar to the source domain, whereas high-level attention masks changed adaptively. In conclusion, the proposed 4D model with attention module performed well and facilitated interpretation of DNNs, which is helpful for subsequent research.

## 1. Introduction

For many years, decoding the brain's activities has been one of the major topics in neuroscience. Inferring brain states consists of predicting the tasks subjects performed and identifying brain regions related to the specific cognitive functions(Friston et al., 1994; Lv et al., 2015; McKeown et al., 1998; Norman et al., 2006). Deep learning (DL) methods based on a variety of artificial neural networks have gained considerable attention in the scientific community for more than a decade, breaking benchmark records in several domains, including vision, speech, and natural language processing(Krizhevsky et al., 2017; LeCun et al., 2015). In this context, deep neural networks (DNNs), especially convolutional neural networks (CNNs), have been recruited for brain decoding(Huang et al., 2018; Li & Fan, 2018; Wang et al., 2019; Yin et al., 2020; Zhang et al., 2021), and achieved high accuracy (> 90%) in brain multiple state decoding(Nguyen et al., 2020; Wang et al., 2020). It is important to note, however, several open challenges still need to be addressed while using deep learning to investigate functional magnetic resonance imaging (fMRI) data.

The first challenge is the abstraction of complex temporo-spatial features within the fMRI time series. A fMRI time series is a four-dimensional (4D) data that consists of three-dimensional (3D) spatial and one-dimensional (1D) temporal information, which means brain regions engage and disengage in time during coherent cognitive activity(Chen et al., 2019; Shine et al., 2016). Inspired by this, Mao et al. (2019) developed a model of 3D CNN stacks and a Long Short-Term Memory (LSTM) for spatial and temporal feature abstraction, respectively. A bit more reasonable approach would be to jointly leverage the inherent spatial–temporal information in fMRI data (Ismail Fawaz et al., 2019). However, designing and optimizing architectures for 4D fMRI decoding is difficult due to the lack of systematic comparisons of various spatiotemporal processing and the substantial explosion of computational and memory requirements.

The second challenge is the researchers' requirement for a higher degree of accountability of the model, which is the core of the feasibility and reproducibility of brain decoding (Lindsay, 2020). Deep learning is regarded as a black-box model, and recent efforts have been made to develop an interpretable brain decoding model through feature ranking (Li & Fan, 2019), visualizing the convolutional kernels (Vu et al., 2020), guided back-propagation(Wang et al., 2020), and so on. Improved CNN interpretability in fMRI analysis could lead to more accountable usage, better algorithm maintenance and improvement, and more open science (Tjoa & Guan, 2021).

Another challenge is the conflict between the DNNs' requirement for large amounts of data and the relatively modest quantity of datasets in typical cognitive research (Yotsutsuji et al., 2021). Most fMRI experiments comprise tens to hundreds of participants due to experimental costs or participant selection. It is natural to use transfer learning to alleviate the data scarcity problem in the target domain (e.g., small sample datasets) by utilizing the knowledge acquired in the source domain (e.g., large cohorts)(Gao et al., 2019; Svanera et al., 2019; Thomas et al., 2019; Wang et al., 2020). fMRI data varies across datasets (e.g. scanner, scanning parameters, task design, template space), so it remains an open question how far the DNN can transfer-learn in fMRI.

Inspired by these challenges, the main contributions to this paper are threefold. First, we extended the problem of temporal modeling and spatial feature extraction to the 4D convolution module and compared various approaches to fMRI data processing. Second, we employed the mixed attention modules to improve the decoding performance, which not only enhanced the ability to distinguish and focus on specific features but also presented an in-depth interpretation of CNN. Third, we explored the benefits of transfer learning in fMRI analysis under different problem definitions and task design, demonstrating that cognitive similarities can extend to individual trait differences.

## 2. Materials and Methods

## 2.1. Dataset

### 2.1.1. Human Connectome Project (HCP) dataset

The minimally preprocessed 3T data from the S1200 release of the HCP(Glasser et al., 2013) was used in this research. The present study included task fMRI of 1,034 subjects during 7 tasks: emotion, gambling, language, motor, relational, social, and working memory (WM). The 7 task, which lasted for about 20 to 30 frames under different conditions during each block, provided a high degree of brain activation coverage(Barch et al., 2013). Thus, the parameter estimates of the model trained on this dataset contained similarities to multiple cognitive domains and were utilized as the source domain in the transfer learning experiment. The HCP S1200 dataset has been preprocessed with the HCP functional pipeline and normalized to the Montreal Neurological Institute's (MNI) 152 space. According to the previous studies(Nguyen et al., 2020; Wang et al., 2020), only one condition was selected for each task (Table 1) and resulted in 14821 fMRI 4D instances across all subjects and tasks. To save the memory of computing, a bounding box of the size of [80,96,88] voxels was applied to each fMRI volume and the blank parts that did not contain brain tissues were cropped out.

**Table 1. Details of the selected HCP time series**

| Task | Selected condition | Frames of the block |
|---|---|---|
| Emotion | fear | 26 |
| Gambling | loss | 39 |
| Language | present story | 29 |
| Motor | right hand | 17 |
| Relational | relational | 23 |
| Social | mental | 32 |
| Working memory (WM) | 2-back places | 39 |

### 2.1.2. BOLD5000 dataset

The BOLD5000 (Chang et al., 2019) dataset was also used for transfer learning of the proposed model. The dataset selected event-related design paradigms to investigate visual perception, which collected fMRI of 4 participants while viewing 5,000 real-world images. Each image was presented for 1 sec and followed by a 9 sec blank screen with a fixation cross. Thus, a single trial lasted 5 frames (Repetition Time, TR = 2s). Two conditions of stimulus images were employed in this study: Scene containing whole scenes and ImageNet focusing on a single object. Implicit image attributes can provide category selectivity in high-level visual regions. Using fMRIPrep (Esteban et al., 2017), the preprocessing including motion correction, distortion correction, and co-registration to the corresponding T1w of the fMRI data was applied. Then each volume was also cropped to the size of [80,96,88] voxels. As a result of using an event-related design, each segmented fMRI input covered the entire trial and included 2 extra TRs extended forward and backward. Thus, the size of the input data was [80, 96, 88, 7].

## 2.2. The Proposed Neural Network

The proposed model consists of a 4D convolution layer and four 3D attention modules, followed by a fully-connected layer (Figure 1a).

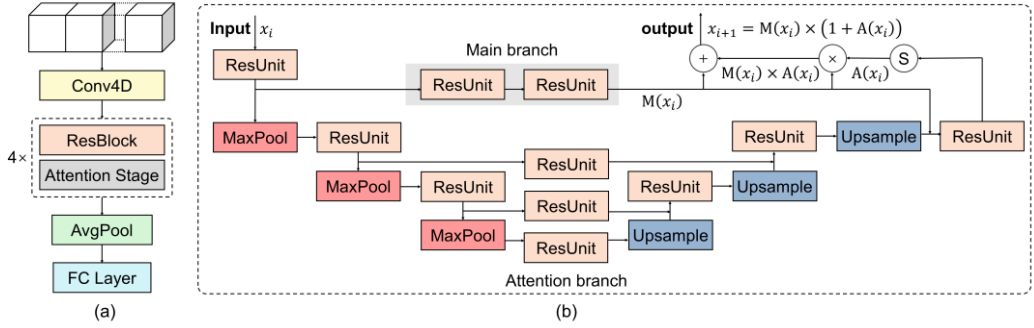

**Figure 1 The proposed neural network.**

(a) The model consists of a 4D convolution layer, four 3D Attention Modules, and a fully-connected layer to provide labeled task classes. (b) The Attention Module, which includes the main branch and an attention branch composed of downsample and upsample paths connected by a shortcut branch.

### 2.2.1. 4D Convolution

The 4D convolution kernel $K \in \mathbb{R}^{k_l \times k_h \times k_w \times k_d \times k_c}$ was applied to the input $x \in \mathbb{R}^{l \times h \times w \times d \times c}$, where $l$ is the temporal length, $h$ is the height, $w$ is the width, $d$ is the depth, and $c$ is the length of the channels. The 4D convolution operation, Conv4D, was implemented by two loops of the native 3D convolution operation, Conv3D, of the Pytorch (Paszke et al., 2019) :

$$(K * x) = \sum_i^{k_l} \sum_j^{(l-k_l)/s_t \ +1} Conv3D(K(i), x(j \cdot s_t + i)) \ ,$$

where $s_t$ is the temporal strides ($s_t$=1, 2, …) and $s_s$ is the spatial strides ($s_s$ =2). A stride of >1 leads to a down-sample in the designated dimension. After the 4D convolution, the temporal dimension was squeezed and flattened to channel dimension before the attention module.

### 2.2.2. Attention Module

The attention mechanism in the DNN selects focused regions and thus enhances the discriminative representation of objects (Vaswani et al., 2017). The attention module is also beneficial for optimizing by serving as a gradient update filter to prevent gradients from noisy regions. Inspired by previous researches(Wang et al., 2017; Woo et al., 2018), we developed a 3D Mixed Attention Module (Figure 1b), where the processing flow was split into the main branch and the attention branch. The main branch serves for feature extraction and retains effective backpropagation. The feature processing in the main branch may be any convolution network structure, and the ResNet block (He et al., 2016) was used in the present work. Formally, the output of the main branch is denoted as M($x$) with an input feature map $x$. The attention branch is a U-shaped architecture (Ronneberger et al., 2015) to mimic the feedforward and feedback attention processes. The downsample path is built by several stacks of a 3D MaxPoll and a ResUnit to capture valuable context from multiple scales. The symmetric upsample path consists of the same amount of trilinear interpolation and ResUnit (Figure 1b). Finally, the output was normalized by a Sigmoid function to obtain the A($x$).

Naive dot production of two branches degrades the value of features. Attention residual learning is used to ease this problem by constructing the attention branch as an identical mapping. Formally, the output of Attention Module $x_{i+1}$ serving as the input of the next layer is modified as:

$$x_{i+1} = M(x_i) \times \left(1 + A(x_i)\right)$$

What's more, the attention mask branch can be viewed as an identical mapping that changes adaptively as layers go deeper. What the neural network learns at each level can be demonstrated by the distribution of attention. The attention masks of each channel were visualized to present an in-depth interpretation of the network by upsampling the feature map corresponding to A($x$) and mapping it to T1w.

## 2.3. Training and Evaluation

The implementation of the different model variants is based on the PyTorch framework. Training was performed on an NVIDIA GTX 1080Ti graphic card. To conduct afair comparison, the batch size was set to 16 and each model was trained for 60 epochs using the Adam algorithm with the standard parameters ($\beta_1$= 0.9 and $\beta_2$= 0.999). The learning rate was initialized at 0.0001 and decayed by a factor of 5 when the validation loss plateaued after 15 epochs. The loss converged well and overfitting was not observed during validation experiments. Our validation strategy employed a fivefold cross-validation across subjects and the dataset was categorized into subsets as follows: training set (70%), validating set (10%), and testing set (20%). Control experiments were conducted on various model variants (Table 2) to verify whether the 4D convolution and attention modules brought a substantial improvement. We also analyzed a set of 4DResNet consisting of different sizes of 4D kernels and presented comparison results using different frames as input. During the testing stages, the performance was evaluated by voting on the output of all segmentations from one data instance.

## 2.4. Transfer Learning

Transfer learning describes a process in which a network is trained on a source dataset and subsequently reuses the parameters of the pre-trained network that contained knowledge about the source domain on the target dataset. Transferability is an important advantage of deep learning methods compared with traditional methods in fMRI decoding. To this end, the transfer learning strategy was applied to evaluate the general use representation of the trained model.

Inter-task (same dataset, different task) transfers. Since fluid intelligence (gF) measures the intelligence-related score which reflects inherent cognitive ability, there is great interest in inferring gF from fMRI data(Greene et al., 2018; He et al., 2020; He et al., 2018). In the HCP data set, gF was quantified using a 24-item version of the Penn Progressive Matrices test. We employed fMRI of working memory (WM) task and split the subjects same as the 7 tasks classification. The transfer training is similar to the initial training, except that the parameters of the low-level layers were pre-trained and the fully connected layers were redefined and initialized. Besides, the loss function was changed to MSE Loss. Thus, we evaluate the performance of transferability by comparing the Spearman's correlation coefficient between the predicted gF and the observed gF of the initial model, the transferred model, and the previous work(Greene et al., 2018).

Inter-datasets (different dataset, different task) transfers. BOLD5000, which selected event-related design paradigms, is another small sample target dataset including 4 participants. The source and target datasets are different in data statistics and distributions. The key idea of this workflow is similar to that mentioned above. We fine-tuned the model to decode different types of stimulus images seen by subjects and randomly used the data from three subjects for training and one subject for testing.

## 3. Results

### 3.1. Performance Evaluation

The performance of various models was compared by the mean and standard deviation of accuracy (Table 2). All of the proposed models effectively distinguished 7 tasks, with the 4DResNet-Att obviously outperforming the others with an accuracy of 97.4%±0.4% (mean ± SD). Figure 2a shows the decoding performance of 4DResNet-Att on 7 cognitive tasks, and the confusion matrix shows a nice block diagonal architecture. The cognitive tasks were accurately identified with the accuracy of: Emotion (96.2 ± 0.2%), Gambling (99.4 ± 0.3%), Language (98.7 ± 0.4%), motor (96.0 ± 0.4%), Relational (93.6 ± 0.9%), Social (99.4 ± 0.3%), and WM (98.9 ± 0.4%). Furthermore, the confusion matrix showed misclassifications of the Relational and the Gambling, the Emotion and the Gambling, the Motor and the Gambling, and the Relational and the WM.

**Table 2. Comparisons with previous methods on HCP dataset**

| Authors | Model | Accuracy± SD |
|---------|-------|--------------|
| Wang et al. (2020) | 3DResNet | 93.7±1.9% |
| Nguyen et al. (2020) | 3DResNet-TF | 95.1±0.6% |
| | 3DResNet-LSTM++ | 97.0±0.4% |
| | 3DResNet-TF++ | 97.2±0.6% |
| Ours | 3DResNet-Att | 96.3±1.1% |
| | 4DResNet | 96.1±0.8% |
| | **4DResNet-Att** | **97.4±0.4%** |

The superior performance of the 4DResNet-Att model in comparison to the 3DResNet(Wang et al., 2020) and other recent researchers (Nguyen et al., 2020) is possibly due to the capability to handle complex spatio-temporal dynamics in fMRI series by 4D convolution operations and the use of the attention mechanism to adaptively select a focused location.

Specifically, the 4DResNet is able to capture dynamic changes in hemodynamic response on temporal dimension and to integrate these representations from interconnected brain regions on spatial dimension. To evaluate whether 4DCNN brings a substantial improvement over 3DCNN, the 4DResNet-Att model was compared with the 3DResNet-Att model on the same brain decoding tasks using different lengths of frames as input (Figure 2b). Overall, the 4DResNet substantially enhanced classification performance compared to the 3DResNet, except for the 7 frame condition. The low performance at shorter fMRI input could be caused by two factors: 1) fewer information in short input, especially in series shorter than a hemodynamic response; 2) the 4DResNet tends to measure the relative dynamic change over a long range. Besides, we also evaluated a set of 4DResNet consisting of different sizes of 4D kernels to decode brain activity. Our results revealed that decoders with a short 4D-kernel size achieved lower decoding performance than decoders using a relatively longer 4D-kernel (Figure 2c).

Furthermore, to establish whether the use of attention mechanisms could enhance fMRI decoding, we compared the 4DResNet with attention modules and the naive 4DResNet. Figure 2c showed the results. The 4DResNet-Att outperformed the naive 4DResNet on the HCP dataset under different sizes of 4D kernel. In addition, the 4DResNet-Att network (about 12h) reduced nearly 1/3 of the training time compared with naive 4DResNet (about 19h) while achieving 90% accuracy. As expected, the capability of the attention mechanism to adaptively learn the focused location brings increased performance while reducing training time.

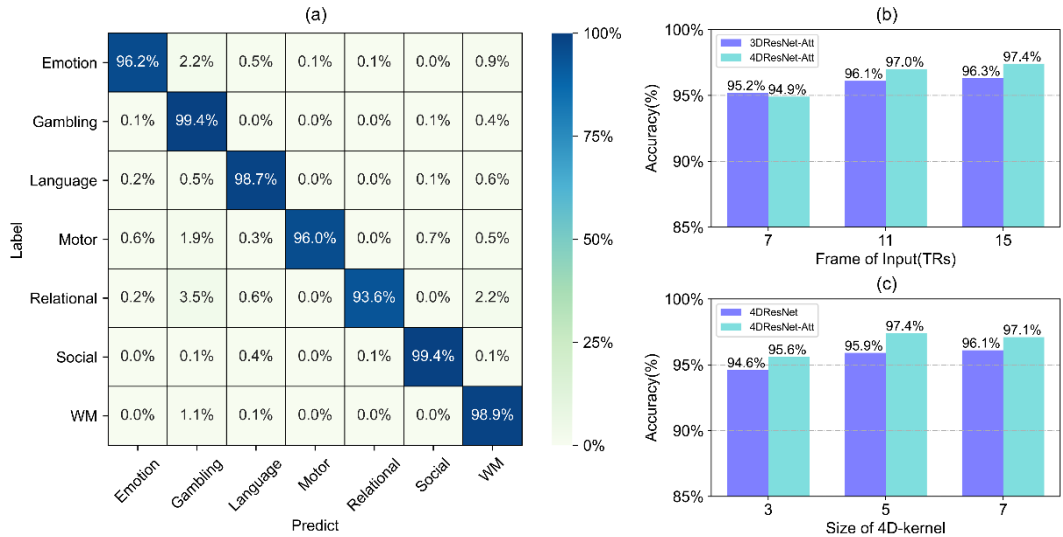

**Figure 2 performance evaluation on HCP dataset**

(a) The average confusion matrix showed a nice block diagonal architecture. (b) The 3DCNN and 4DCNN comparisons used different frames as input (frame = 7,11,15). 4DCNN outperformed in dynamic change over a long range. (c) The classification performance with or without the attention module (frame = 15). Decoders with attention and a relatively longer 4D-kernel performed better.

## 3.2. Visualization of Attention Mask

Previous studies have employed some visualizations to build an interpretable brain decoding model in fMRI analysis(Vu et al., 2020; Wang et al., 2020; Yin et al., 2020). Here, we visualized the focused regions of the attention module in each convolution layer to present an in-depth interpretation of the

DNN. Each channel obtained seven attention masks for different tasks, which are averaged across all of the input samples from all of the subjects.

Overall, the resulting attention masks from the 1st stage (Figure 3a) have excellent coverage of the brain and prefer to highlight the areas containing the useful BOLD signal, such as the whole brain structure, and diminish the noise areas, such as the background, brainstem or cerebrospinal fluid areas. The masks from the 2nd stage (Figure 3b) have a more concentrated focus than those of the 1st stage, such as the enhancement of gray matter. The attention at this stage focuses on some functional networks and cerebral cortex related to different cognitive functions, such as the default mode network, sensorimotor network, temporal cortex, and visual cortex. The attention mask from the 3rd stage is even more focused, covering task-specific brain areas (Figure 3c). It is notable, however, the focused layouts of the attention mask varied across different tasks and were remarkably task-specific. A channel could generate different weights for specific areas for each of the different tasks, such as the left motor cortex areas for motor task, both superior and inferior temporal cortices for language task, and the anterior prefrontal cortex for relational task. The attention mask from the 4th stage is coarser than those of low levels due to the stride in the convolution operation (Figure 3d). What's more, the weight of the attention mask has a narrower range, which could be due to the fact that the corresponding feature map of the main branch is abstract. The attention masks also serve as a gradient update filter, so a small range of attention weight in the high level feature map could prevent some gradient problems.

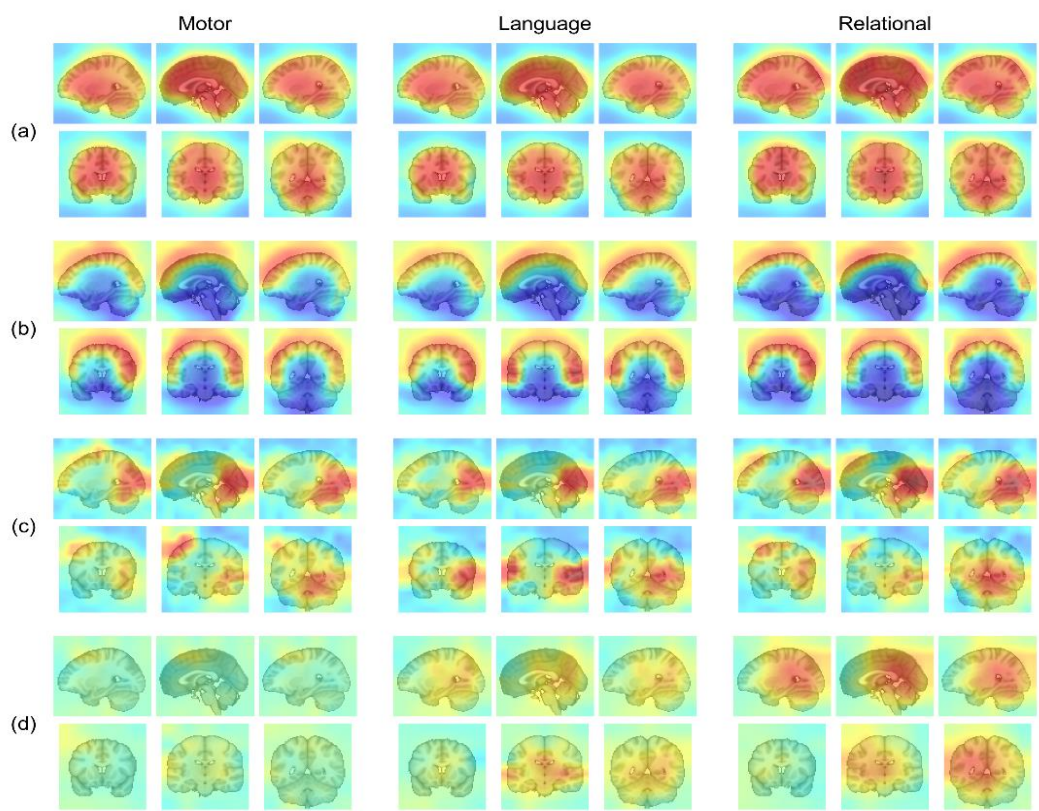

**Figure 3 Visualization of attention mask**

The average focused regions on four attention stages (a-d: from bottom to top) of different tasks

(Language, Motor, Relational). Each of the attention masks was color-coded with a color gradient indicating the enhancement (positive) or diminishment (negative) of the feature maps.

### 3.3. Transfer Learning

Here, two different approaches were used to explore the benefits of transfer learning in fMRI analysis under different problem definitions or task design.

First, we evaluated the general use of representation of the trained model between different problems, from cognitive similarities of group to individual trait differences in subjects. Recent research has demonstrated that connectome-based predictive modeling built from task-based fMRI data improves prediction of individual traits(Greene et al., 2018). Here, the knowledge about similarities and differences between intrinsic and task-induced brain states contained in a pre-trained model was transferred to the WM-trans-set, which is a dataset including the WM task, to predict individual trait differences. Figure 4a shows that the transferred regression model yielded significant predictions of gF. The performance of the transfer learning method ($r_s$=0.354, $p$<0.001) evaluated by the average Spearman's correlation coefficient is better than the previous study(Greene et al., 2018) ($r_s$=0.325, $p$=0.001). What's more, the initial model using the same architecture achieved a lower correlation coefficient in prediction ($r_s$=0.306,$p$<0.001). Furthermore, the visualization analysis shows that low-level attention masks (Figure 4b-4c) remained distributed similarly to the source domain, whereas high-level attention masks (Figure 4d-4e) changed adaptively as knowledge transferred from group similarities to individual differences.

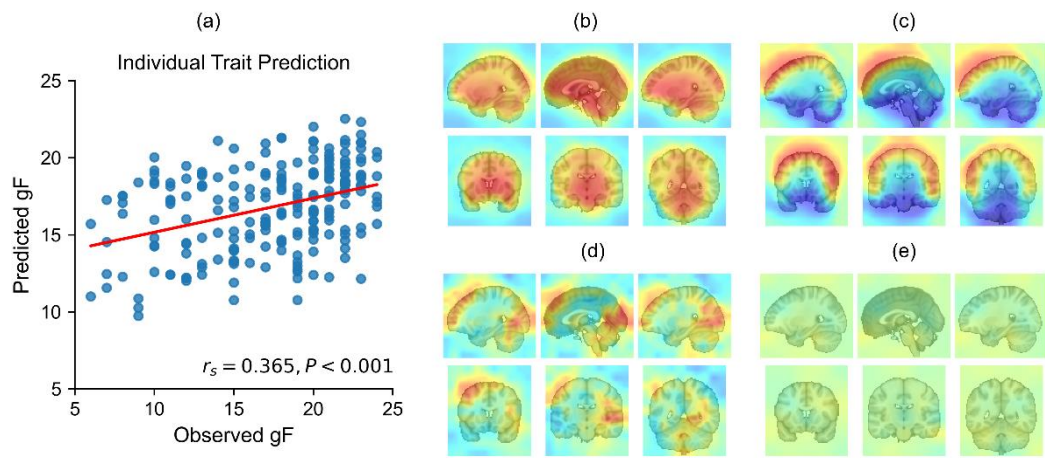

**Figure 4 Individual Trait Prediction**

(a) The transferred regression model yielded significant predictions of gF ($r_s$=0.354, $p$<0.001) (b-e) The attention mask from bottom to top after transfer learning. The focused regions of high levels change adaptively.

Second, the pre-trained model from the HCP dataset was fine-tuned to decode different types of stimulus images on BOLD5000. As can be seen in Figure 5a, the transferred model achieved 77.6% accuracy, which suggests that the knowledge learned from the source domain is highly applicable to the target domain, while the initial model trained from scratch failed to converge to a satisfactory accuracy (<60%) across a wide range of choices of hyper-parameter. Furthermore, the visualizations demonstrated that the attention mask changed adaptively to fit individual subjects' brain structures, despite the fact that the fMRI data was registered to the corresponding T1w space rather than the standard MNI152 space. As the model was fine-tuned to decode visual tasks, the attention masks from the high levels (Figure 5d-5e) also changed adaptively to reweight task-related brain regions.

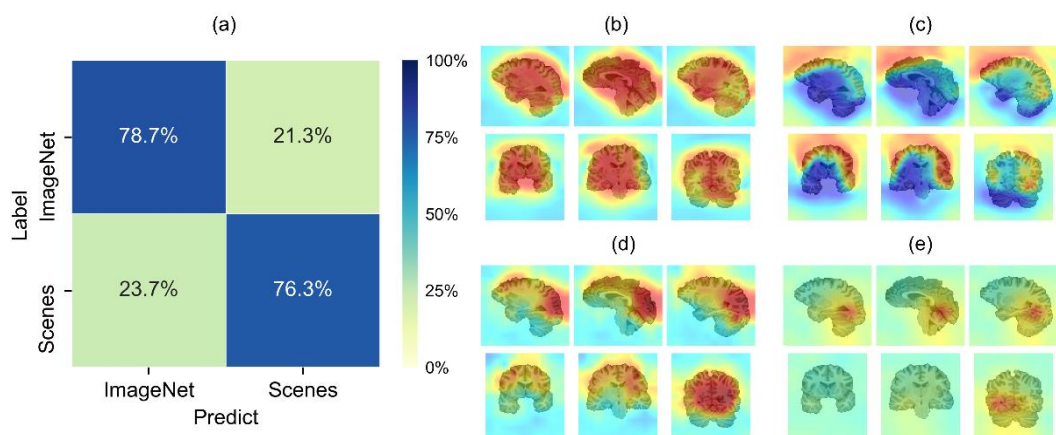

**Figure 5 BOLD5000 visual tasks**

(a) The average confusion matrix on all subjects (accuracy = 77.6%) (b-e) The attention mask from bottom to top after transfer learning. The masks adaptively change to fit different subjects' brain structures.

## 4. Discussion

### 4.1. 4D Convolution

Brain decoding has been a popular topic in neuroscience for decades. Recently, DNNs have gained considerable attention in the scientific community and shown promising performance in brain decoding. The fMRI data is a 4D data consisting of a time series of 3D brain volumes. 4D CNN has shown the feasibility of 4D medical applications, such as 4D computed tomography (CT) (Clark & Badea, 2019) and OCT-based force estimation (Gessert et al., 2020). However, the fMRI data is big and a full 4D DNN is too large to be applied and efficiently trained. Thus,(Wang et al., 2020) proposed a model of 1-D convolution in the first layer for the abstraction of temporal features, followed by stacks of 3D CNNs for spatial features. Mao et al. (2019) developed a network architecture that extracted spatial features out of each fMRI frame using 3D CNNs and passed these latent features to a Long Short-Term Memory (LSTM) network to take into account the temporal dependencies within task-evoked brain activity. The model we proposed includes a 4D convolution layer to detect temporo-spatial features, and puts the features into the channel dimension of the following 3D layers to reduce memory consumption. Above results suggest that the proposed model has a good balance of accuracy and efficiency. Our model could achieve better performance while taking less time than the previous state-of-art works.

## 4.2. Interpretation of CNN

The attention modules not only improve decoding performance but also serve as a visualization tool to investigate how neural networks work in fMRI decoding. The attention mechanism helps humans to mainly focus on the most useful information in the human perception process. Inspired by this, attention mechanisms have been studied extensively in many deep learning fields(Vaswani et al., 2017; Wang et al., 2017; Woo et al., 2018). In this research, the proposed 3D Mixed Attention Module consisted of a main branch and an attention branch and considered both channel and spatial features. The experimental results demonstrate that attention modules have many advantages. For example, the architecture with attention modules was trained to converge faster and more easily and achieve better performance, which could be due to the attention mechanism reweighting the focused areas to enhance discriminative features. The attention module is also beneficial for optimizing during back propagation, which serves as a gradient update filter to prevent gradients from noisy and enhance gradients from important regions.

What's more, cognitive neuroscience research requires a higher degree of accountability, while an end-to-end trainable network has always been regarded as a black-box in neuroscience. Presenting an in-depth interpretation of a method can demonstrate the feasibility and reproducibility of fMRI studies(Li & Fan, 2019; Vu et al., 2020). Visualizing attention distributions allows researchers to investigate what the neural network learns at each level. The low-level masks provide excellent coverage of the brain to highlight useful structures while pruning noisy areas. As the layers go deeper, the attention masks become coarser and tend to focus on various cerebral cortex and functional networks. The high-level attention mask varied across different tasks, re-weighting more attention to the areas related to the specific target task. What's more, the attention mask adapted to fit different subjects' brain structures. This also suggests that our architecture could be a suitable approach to avoid individual variability across subjects in the raw and minimally preprocessed fMRI series without spatial normalization. Besides, the attention areas which could present biologically meaningful interpretations of cognitive neuroscience demonstrated that the proposed CNN decoded states from task-related activations but not from nuisance variables.

## 4.3. Transfer learning

Transferability has been demonstrated to be a significant advantage of DL methods over traditional methods in fMRI decoding(Gao et al., 2019; Wang et al., 2020). To this end, we explored the benefits of transfer learning under various conditions. The transferred regression model yielded significant predictions of individual trait differences and achieved better Spearman's correlation coefficient than the previous study(Greene et al., 2018). This could be due that the previous study relied on the discriminative power of feature selections, and not all connectivity parameters are relevant for prediction, while the transferred model could automatically capture the full range of individual trait differences. This also suggests that the group cognitive similarities among intrinsic brain states could generally be reused to predict individual differences, which is important for precision medicine in clinical research. Furthermore, previous studies most commonly applied transfer learning between the block-design dataset. On the BOLD5000, the pre-trained model from the HCP dataset was fine-tuned to decode different visual tasks and obtained 77.6 %. Despite the fact that the model was trained using the block-design dataset, the internal properties of human hemodynamic responses contained in the parameters are consistent and could be reused in the event-design dataset.

## 4.4. Limitations and Future Applications

In this project, the proposed model outperformed other architectures. Despite the 4D convolution processing dynamic changes more efficiently, some limits remain, such as a substantial increase in computational and memory requirements. What's more, we only chose one condition for each cognitive domain in order to be comparable to previous studies, while the BOLD signals might be a mixture of hemodynamic responses evoked by different task events. A decoding model with fine cognitive granularity would generalize similarities and differences among task-induced brain states from multiple cognitive domains, which is important for transfer learning. The visualization result demonstrated that the high decoding performance was driven by the response of biologically meaningful brain regions. However, the statistical property of the attention mask remains unclear. We could have the results of qualitative analysis and should be cautious until further investigations into its reliability and statistical properties. The transfer learning method, which successfully extended similarities in brain activity to individual differences, showed potential for research in psychiatry and neurology. The pre-trained model based on cognitive state can serve as a brain information retrieval system to distinguish differences in neurologic diseases and classify different psychiatric categories.

## 5. Conclusion

In this study, we designed a 4DResNet with attention module for brain decoding. After investigating the efficacy of some alternative classifiers, the proposed 4DResNet-Att achieved 97.4% on the HCP dataset. We further demonstrated the model's transferability to a variety of tasks and datasets and presented an in-depth interpretation of the network. The visualization analysis of attention distributions illustrated the hierarchical emergence of task-specific masks with depth. After transfer learning, the adaptively changed attention distribution demonstrated the representation could be general extended from cognitive similarities to individual differences.

## Declaration of Competing Interest

The authors declare that they have no known competing financial interests or personal relationships that could have appeared to influence the work reported in this paper.

## Acknowledgments

This work was supported by the National Natural Science Foundation of China (grant nos. 81701665, 21876041), the Fundamental Research Funds for the Central Universities (WK5290000002)